\documentclass[%
 reprint,
 longbibliography,
superscriptaddress,
showkeys,
 amsmath,amssymb,
 aps,
prl,
twocolumn,
floatfix,
floatfix,
]{revtex4-1}

\usepackage[utf8]{inputenc}
\usepackage{graphicx}
\usepackage{dcolumn}
\usepackage{bm}
\usepackage{array}

\usepackage{siunitx}
\usepackage{amssymb}
\usepackage{algpseudocode} 
\usepackage{algorithm}
\usepackage{amsmath}

\usepackage{hyperref,color}
\definecolor{linkColor}{rgb}{1,0,0}
\hypersetup{pdfborder={0 0 0},colorlinks=true,urlcolor=linkColor,citecolor=linkColor}
\newcommand{ \degree }{^\circ}

\begin{document}
\preprint{APS/123-QED}

\title{A Practical Reconstruction Method \\ for \\ Three-Dimensional Phase Contrast Atomic Electron Tomography}


\author{David Ren}
\email{david.ren@berkeley.edu }
\affiliation{Department of Electrical Engineering and Computer Sciences, University of California, Berkeley, CA 94720, USA}

\author{Michael Chen}
\affiliation{Department of Electrical Engineering and Computer Sciences, University of California, Berkeley, CA 94720, USA}

\author{Laura Waller}
\affiliation{Department of Electrical Engineering and Computer Sciences, University of California, Berkeley, CA 94720, USA}

\author{Colin Ophus}
\email{cophus@gmail.com}
\affiliation{National Center for Electron Microscopy,
Molecular Foundry,
Lawrence Berkeley National Laboratory, 
1 Cyclotron Road,
Berkeley, CA 94720, USA}

\date{\today}
\begin{abstract}

Electron tomography is a technique used in both materials science and structural biology to image features well below optical resolution limit. In this work, we present a new algorithm for reconstructing the three-dimensional(3D) electrostatic potential of a sample at atomic resolution from phase contrast imaging using high-resolution transmission electron microscopy. Our method accounts for dynamical and strong phase scattering, providing more accurate results with much lower electron doses than those current atomic electron tomography experiments. We test our algorithm using simulated images of a synthetic needle geometry dataset composed of an amorphous silicon dioxide shell around a silicon core. Our results show that, for a wide range of experimental parameters, we can accurately determine both atomic positions and species, and also identify vacancies even for light elements such as silicon and disordered materials such as amorphous silicon dioxide and also identify vacancies.


\end{abstract}

\maketitle

\section{Introduction}
\label{sec:intro}

Transmission electron microscopy (TEM) offers various imaging modes, allowing for quantitative 3D estimations of local structure, electrostatic and magnetic potentials, and local chemistry~\cite{midgley2009electron}, providing significant impact in both biology and materials science~\cite{luvcic2013cryo, leary2012recent}. It is now possible to measure the 3D position of individual atoms with high precision~\cite{van2011three, bals2011three, xu2015three}, and even determine both the 3D postion and species of every atom in a nanoscale sample with high reliability~\cite{Yang:17}. These atomic electron tomography (AET) studies used a TEM imaging mode called annular dark field (ADF) scanning transmission electron microscopy (STEM). ADF-STEM imaging offers monotonic contrast that is close to a linear 2D projection of the 3D electrostatic potential of the sample. This feature allows for both traditional tomographic reconstruction algorithms~\cite{friedrich2005comparison} and advanced algorithms that allow for some deviation from linearity~\cite{pryor2017genfire}. However, this imaging mode requires large electron doses, as it is much less efficient than phase contrast imaging modes~\cite{ophus2016efficient, chen2016line}. Additionally, because the electron probe is focused to a small spot and scanned over the sample surface, sample motion during the experiment can cause artifacts~\cite{ophus2016correcting}. 

The simplest phase contrast imaging mode used in atomic resolution TEM studies is plane-wave illumination, usually referred to as high resolution transmission electron microscopy (HRTEM). However at atomic resolution, HRTEM imaging produces highly nonlinear contrast for any sample thicker than a few atomic monolayers, often requiring image simulation to interpret the results~\cite{Kirkland:2010, urban2008studying}. For thin samples, comparing experiments to simulations can recover some quantitative 3D information~\cite{argentero2017unraveling}, but this is difficult or impossible for experiments with a high degree of multiple electron scattering. Thus, phase contrast imaging is not widely used in materials science electron tomography studies at atomic resolution.

By comparison, phase contrast HRTEM imaging in biology is simpler to interpret because most biological specimens can be approximated as ``weak phase objects,'' allowing for the phase contrast induced by the sample to be reconstructed from a single defocused intensity measurement \cite{downing2008restoration}. This single-image requirement is important because most biological samples are extremely sensitive to electron beam damage and cannot tolerate high electron doses without damage, orders of magnitude lower than those typically used in materials science \cite{egerton2004radiation}. In structural biology, the introduction of direct electron detectors with high quantum efficiency \cite{mcmullan2014comparison} has rapidly expanded the number of solved protein structures, using 3D tomographic averaging of images of many identical or near-identical protein structures with random orientations.  This technique is called single particle cryo-electron microscopy (cryo-EM) \cite{doerr2015single}. When imaging larger biological samples, averaging of identical sub-volumes can also produce very high resolution reconstructions \cite{briggs2013structural}. 

\begin{figure*}[!thbp]
	\centering
	\includegraphics[width=6.0in]{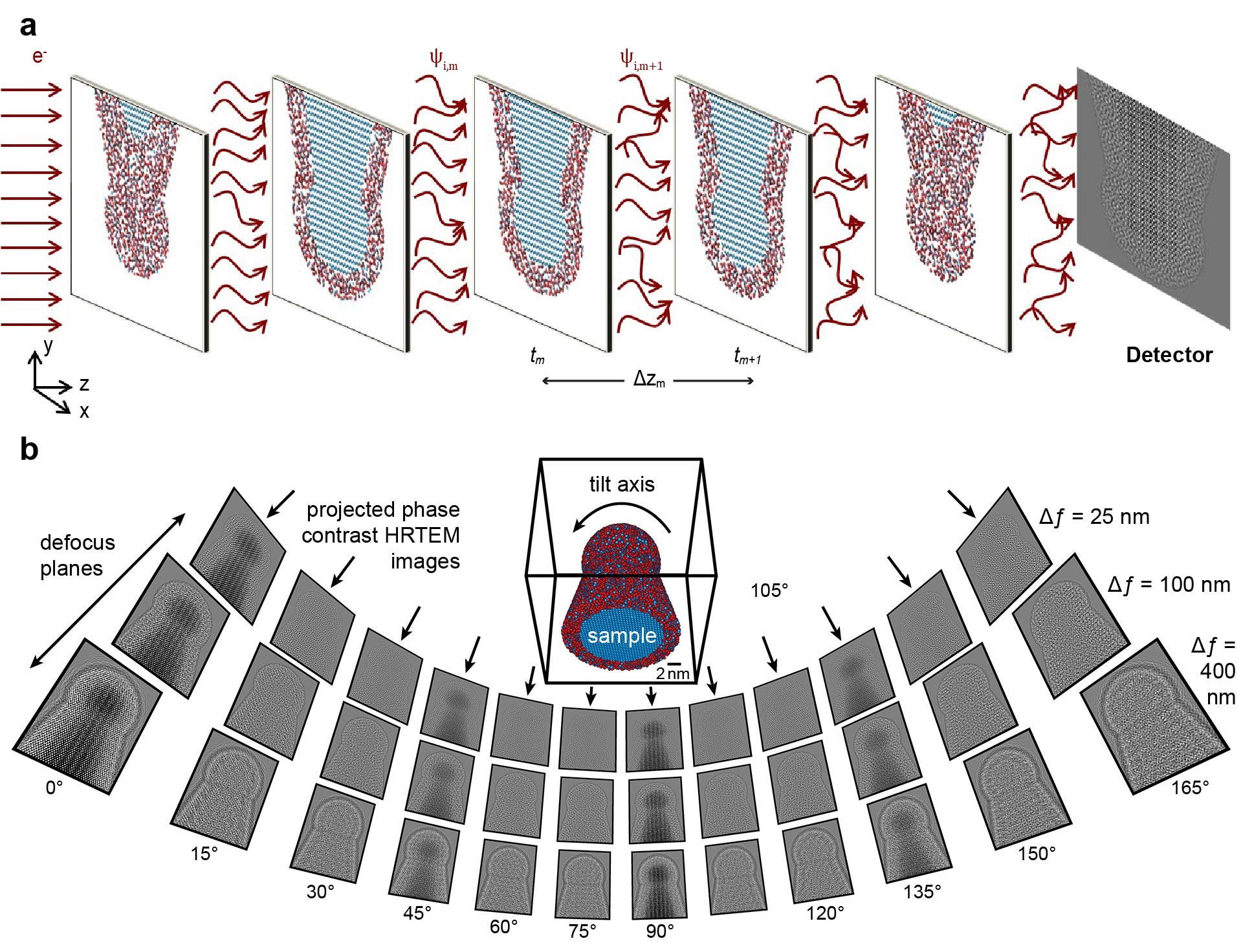}
	\caption{Experimental layout of a phase contrast atomic electron tomography experiment for a core-shell SiO$_2$ with a needle geometry. (a) Multislice forward model that accounts for multiple scattering events. The sample is tilted at an angle and illuminated by on-axis plane wave. (b) The sample is tilted with respect to the electron beam, and at each tilt angle one or more defocused HRTEM images are recorded, up to a tilt range of $180\degree$.}
	\label{FigureSchematicGeometry}
\end{figure*}

Recent advances in computational methods have improved reconstruction accuracy even further, for example by introducing a three-dimensional correction of the microscope contrast transfer function (CTF) \cite{bharat2015advances}. These methods usually assume that the sample can be modeled as weak phase object, and that it satisfies the projection assumption (i.e. measured signal is a linear sum of the projected potential)\cite{Vulovic:2014}, in order to linearize the physical model to derive a closed form solution. However, these assumptions usually only hold for very thin samples~\cite{ophus2017automatic}. Nonlinear effects such as multiple scattering are no longer negligible for thick samples, which represent a large majority of materials science samples at atomic resolution. Therefore, moving beyond these assumptions requires both nonlinear forward model as well as reconstruction method that captures the dynamical scattering of the electron beam.


In general, the interaction of the electron beam with a sample can be modeled with two linear operators: The first is the specimen transmittance function multiplication, which models the absorption and phase delay of the electron beam when interacting with the sample. The second is the Fresnel propagation operator, where the beam moves through free space to the next region of the sample \cite{Kirkland:2010}. Unfortunately, these two operators do not commute, making the inverse scattering calculation both nonlinear and non-convex. Van den Broek and Koch have proposed an inversion method for multiple electron scattering, which uses multiple tilt projections (of relatively small angles) for phase contrast TEM imaging to perform a 2+1 dimensional reconstruction \cite{VandenBroek2012method,VandenBroek:2013}. They have demonstrated in simulation the possibility of reconstruction of the atomic potential of a small nanoparticle in 3D from a small number of tilt angles, for strongly scattering atoms and a low TEM accelerating voltage of 40 kV. Alternative methods to correct for multiple scattering or to directly interpret complex interference patterns in order to perform 3D phase contrast reconstructions have been proposed in optics\cite{maiden2012ptychographic, suzuki2014high, kamilov2015learning, waller2015computational, gao2017electron, samelsohn2017invertible, li2018multi}. Particularly in reference \cite{Tian:15:3D}, the authors demonstrated the recovery of 3D indices of refraction from intensity-only images captured by coherently illuminating the fixed sample from multiple angles. 

In this paper, we present a new method for 3D tomographic reconstruction from a tilt series of one or more intensity-only images at different defocus values.  Our algorithm models the multiple scattering of the electron beam and the strong phase shifts induced by individual atoms at atomic resolution, and includes efficient regularization to recover a physically accurate structure even with very low signal to noise ratio(SNR). We have also implemented an atom-tracing algorithm that is capable of identifying individual atoms as well as estimating their sub-voxel 3D positions and chemical species. Our proposed method will allow for AET experiments to be performed on samples that contain weakly scattering elements such as carbon, oxygen or even lithium, with either crystalline or amorphous structures, or a mix of both. Additionally it will allow for reconstruction of samples that cannot withstand the required electron dose for existing atomic resolution reconstruction methods. Biological cryo-EM studies may also benefit if they are performed on very large volumes (where the projection assumption breaks down) or contain multiple scattering regions. Finally, we are releasing all source codes online along with detailed instructions in order to facilitate widespread adoption of our method.



\section{Methods}
\label{sec:methods}

\subsection{Atomic Structure}

\begin{figure}[!htbp]
	\centering
	\includegraphics[width=3.3in]{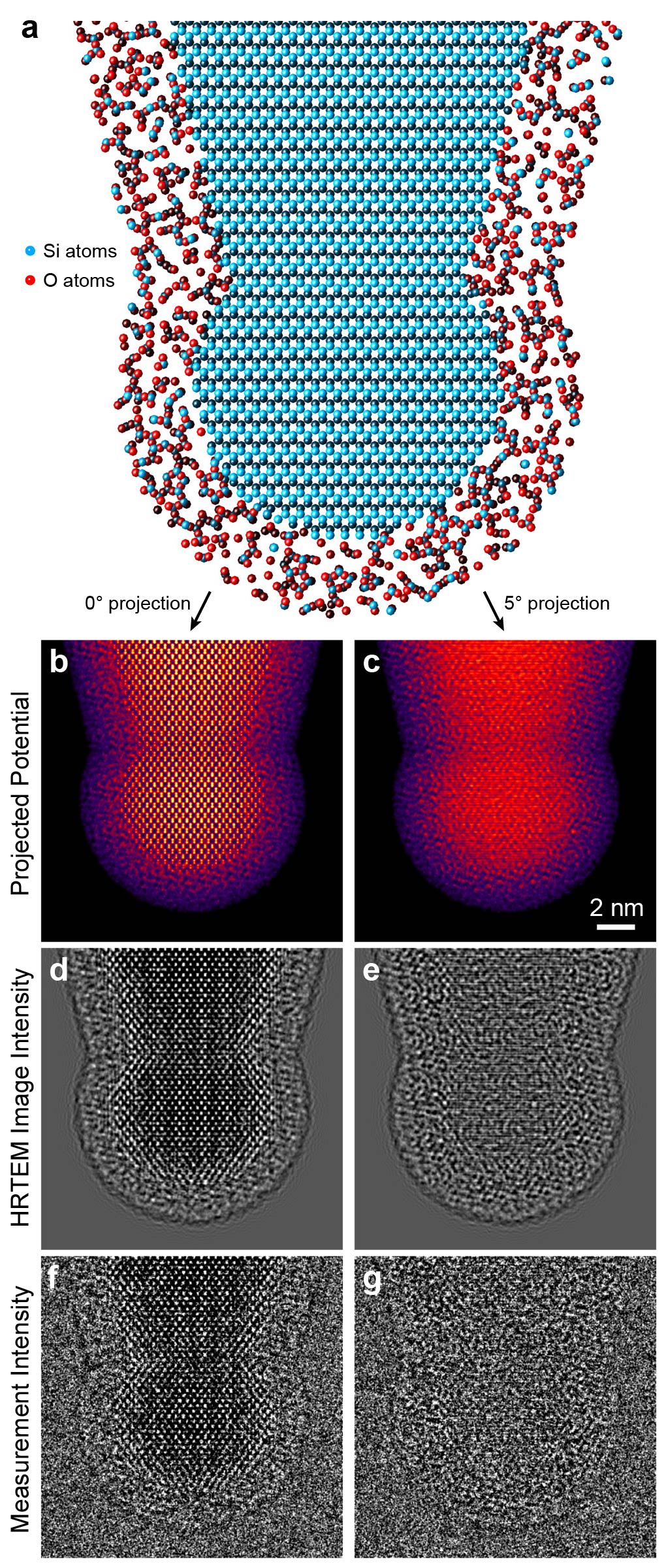}
	\caption{Forward HRTEM simulation of the SiO$_2$ model. (a) A slice of the atomic structure, perpendicular to the electron beam direction. The summed 2D projected potential of the object after (b) $0\degree$ and (c) $5\degree$ rotation, with intensity scaled to show the weakly scattering edges. (d),(e) The ideal HRTEM image at 100 nm defocus for (b) and (c) respectively, and (f),(g) the same images with a dose of 40 electrons / $\mbox{\AA}^2$.}
	\label{FigureSchematicSimulation}
\end{figure}

In this study, we consider a two-component sample structure, with a needle or tip geometry (similar to the experiment described in \cite{xu2015three} and shown in Figure \ref{FigureSchematicGeometry}). The structure used consists of a crystalline silicon core, and a silicon dioxide outer layer. The crystalline core has a tip diameter of approximately 10 nm, which has been achieved experimentally \cite{swanwick2014nanostructured}. A 2 nm thick shell of SiO$_2$ was placed over the entire Si tip. The SiO$_2$ coordinates were taken from the SiO$_2$ structure given in \cite{zhu2013towards}, computed using Density Functional Theory (DFT). This structure was randomly rotated and tiled, before cropping on both the inner and outer surfaces. Finally, a 1.2 \AA\,  minimum distance was enforced between the atomic positions of the Si core and SiO$_2$ shell. In total 150 847 atoms are present in the structure used here. A slice of the atomic coordinates are plotted in Figure \ref{FigureSchematicSimulation}(a), showing the core-shell structure.

The overall structure of this sample is complex, containing both fully crystalline and fully amorphous regions which are stacked along the beam direction for all projection directions. Additionally, while silicon scatters the electron beam with a moderate cross-section, oxygen atoms scatter only weakly. And finally, the amorphous SiO$_2$ structure has an Si-O bond length of approximately 1.6 $\rm{\AA}$~\cite{devine1987si}, making it challenging to resolve the individual atoms in this structure. These attributes make this structure a good test of practical AET reconstruction algorithms for challenging samples, at length scales that can be achieved in experimental samples.

\subsection{Experimental Geometry}

The imaging method we consider is the simplest TEM measurement protocol: plane-wave illumination, typically referred to as HRTEM, or phase contrast imaging. Using a modern TEM instrument equipped with a hardware aberration corrector, we can image the sample with very low amounts of coherent wave aberrations present in the electron beam, with sufficient coherence for atomic resolution imaging \cite{haider1998electron, batson2002sub}. Figure \ref{FigureSchematicGeometry}(b) shows examples of HRTEM plane wave images.

In this study we consider the effect of defocus on the HRTEM imaging. Figure \ref{FigureSchematicGeometry}(b) shows the effect of defocusing the electron wave, which is to increase the contrast and delocalize the atomic signal. In this near-field, or Fresnel diffraction regime, each image is high-pass filtered by the microscope, and the measured signal is modulated by the CTF~\cite{Kirkland:2010}. This CTF can lead to one or more pass-bands in the image spatial frequencies, as well as contrast inversions. 

The sample is mounted on a tilt-rotation stage. This allows it to be rotated with respect to the electron beam. In typical electron tomography experiments, the sample can be tilted over a large range of angles. For the needle geometry considered here, a full tilt range of $180\degree$ has been demonstrated \cite{xu2015three}, using the tilt-rotation piezoelectric TEAM stage \cite{dahmen2009background}. However, most electron tomography experiments have a ``missing wedge'' of tilt angles where the sample geometry or sample stage prevent measurements of some projection angles. Therefore we also consider the missing wedge issue. In the rest of the paper, we use the terms tilt and rotation interchangeably.

In Figure \ref{FigureSchematicGeometry}(b), strong ``amplitude contrast'' is observed in the tilt directions most closely aligned with the crystalline silicon region of the sample (the low index zone axis imaging conditions). The origin of this amplitude contrast is that a significant portion of the electron beam is scattered to higher angles than the numerical aperture of the experiment by the aligned atomic columns, often called ``channeling'' contrast \cite{humphreys1979scattering}.

\subsection{Microscope and Simulation Parameters}
For our simulation setup, we chose parameters that can be realistically achieved in experiments:

\paragraph{Electron energy:} In order to achieve very high resolution, we use the same electron accelerating voltage as previous AET works~\cite{xu2015three, Yang:17}, which is 300 kV. This energy corresponds to electron beams with de Broglie wavelength of ~0.0197 \AA. While SiO$_2$ is known to be sensitive to the electron beam, it has been imaged using 300 kV HRTEM in past studies \cite{niwa1992sio2, hochella1988reassessment, krumeich2004electron,he2014situ}.

\paragraph{Voxel size:} The voxel size is isotropic in all 3 dimensions. The voxel size of 0.5 \AA\, was chosen because it offers a good balance between resolution, and the field of view / practical computational limits for the size of the reconstructed volume. This voxel size can resolve individual atoms in the amorphous SiO$_2$ structure, which has an average Si-O bond length of 1.6 \AA, as mentioned previously. Due to computational limitations, we use a reconstruction volume of $(24 \rm{nm})^3$, corresponding to $480^3=1.1 \cdot 10^8$ voxels, which requires 422 MB of storage space for each full array at single floating point precision. Because we operate in complex space, the storage size requirements of the volume doubles, and it becomes 844 MB. Without loss of generality, our final reconstruction volume contains a large majority of the sample, which includes approximately 120 000 atoms. 

\paragraph{Tilt angles:} Referring to the Fourier diffraction theorem in the literatures of optical diffraction tomography~\cite{muller2015theory, kakslaney1988principles}, each image measurement in tomography provides information about a particular subspace of the sample's Fourier spectrum (specifically a planar slice). Therefore, the tilt angles are chosen to be equally spaced, in order to best span the information contained in Fourier space. In contrast to crystalline samples, amorphous materials have no preferred measurement directions and therefore are optimally sampled by equally spaced projections \cite{collins2017entropic}. In order to mimic experimental limitations of many tomography studies, we have also examined the effect of a missing wedge where some range of tilt angles are missing.

\paragraph{Defocus:} When few defocus measurements are available, a linear increase of the defocus distance does not optimally collect information about the sample~\cite{Zhong:14}. Instead, for each tilt angle, we mimic an exponential increase in defocus distances. Additionally, for an aberration corrected microscope, positive and negative defocus provide essentially identical information about the sample up to a sign difference. Therefore, we defocus the electron wave in one direction only. We have restricted the defocus to small enough magnitudes to enable easy translation alignment of multiple images, which is required for reconstructions utilizing experimental data. Due to the increased signal delocalization, large defocus values also require a larger field of view and correction of any magnitfication or rotation errors, which would increase the difficulty of the experiment.

\subsection{Forward Simulation}

The forward simulation model that we use is composed of three parts: object rotation, complex wave propagation, and imaging. We describe a 3D object with a series of projected 2D atomic potential functions $V \triangleq \{V_m(\mathbf{r})\}_{m=1}^{N_z}$, where $\mathbf{r}=(x,y)$ are the lateral coordinates and $m$ is the index of the slices along the axial direction ($z$)\cite{Kirkland:2010}. We describe the slice separation by a set $\{\Delta z_m\}_{m=1}^{N_z}$.

First, for each tilt angle $\theta_i$ ($i = 1, 2, ..., N_{\theta}$), we rotate the 3D object along the $y$ axis using the fast rotation algorithm described in~\cite{Paeth:86}. The tilted object $W_i$ is then $W_i = \mathcal{R}_{\theta_i}\left\{V\right\}$, where $\mathcal{R}_{\theta_i}$ denotes a linear rotation operator. 

Then, we model the complex wave, with relativistically corrected electron wavelength $\lambda$, propagation through the object using a multislice algorithm (also known as the beam propagation method), which is able to account for multiple scattering events~\cite{Kirkland:2010, VandenBroek:2013}, as shown in Figure \ref{FigureSchematicGeometry}(a). We first convert each slice from a projected 2D potential function to a 2D transmittance function $t_{i,m}(\mathbf{r}) = \exp\left[i \sigma W_{i,m}(\mathbf{r})\right]$, where $\sigma$ is the beam-sample interaction parameter that depends linearly on $\lambda$. The projected potential of two slices are plotted in Figures \ref{FigureSchematicSimulation}(b) and (c).

The complex electron wave function before reaching each slice is denoted by $\psi_{i,m}(\mathbf{r})$. As it passes through the slice, it will be multiplied by the corresponding 2D transmittance function at the corresponding $z$ depth. After that, it is propagated in free space to the next slice using the angular spectrum:
\begin{eqnarray}
\label{eq:slice_prop}
\psi_{i,m+1}(\mathbf{r}) = \mathcal{P}_{\Delta z_m}\left\{t_{i,m}(\mathbf{r})\psi_{i,m}(\mathbf{r})\right\},
\end{eqnarray}
where 
\begin{equation}
\mathcal{P}_{\Delta z_m}\{\cdot\} = \mathcal{F}^{-1}\left\{\exp\left[i2\pi\Delta{z_m}\sqrt{1/\lambda^2 - \left\|\mathbf{q}\right\|^2}\right]\cdot\mathcal{F}\left\{\cdot\right\}\right\}
\end{equation}
is the linear operator for free-space propagation by distance $\Delta z_m$, $\mathbf{q} = (q_x, q_y)$ is the 2D Fourier space coordinates, and $\mathcal{F}\{\cdot\}$ and $\mathcal{F}^{-1}\{\cdot\}$ denote Fourier transform and its inverse, respectively.

The exit waves of a thin sample (in focus) will show primarily amplitude contrast, but most of the electron scattering information is encoded as phase shifts on the exit wave. In order to create stronger phase contrast, we defocus the exit waves by distances of $\{\Delta f_j\}_{j=1}^{N_f}$ after they pass through the object. The defocus operator is simply another free space propagation, where the propagation distance can be controlled arbitrarily. Then, the intensity of the exit waves are captured:
\begin{equation}
\label{eq:exitwave}
\hat{I}_{i,j}(\mathbf{r}) = \left|\mathcal{H}\left\{\mathcal{P}_{\Delta f_j}\left\{\psi_{i,N_z+1}(\mathbf{r})\right\}\right\}\right|^2 \triangleq \left|\psi_{\rm{exit},i,j}(\mathbf{r})\right|^2,
\end{equation}
where
\begin{equation}
\mathcal{H}\{\cdot\} = \mathcal{F}^{-1}\left\{H(\mathbf{q})\cdot\mathcal{F}\left\{\cdot\right\}\right\},
\end{equation}
with $H(\mathbf{q})$ denoting the transfer function of the microscope, usually called the contrast transfer function (CTF) \cite{Kirkland:2010}. After all tilt angles and defocus images are acquired, we obtain a series of images $\{\hat{I}_{i,j}(\mathbf{r})\}_{i=1,j=1}^{N_{\theta},N_f}$, examples of which are shown in Figures \ref{FigureSchematicSimulation}(d) and (e). The multislice beam propagation method is outlined in Algorithm \ref{alg:forward}, and the schematics are shown in Figure \ref{FigureSchematicGeometry}(a).
\begin{algorithm}[H]
    \caption{Forward model computation}
    \label{alg:forward}
    \textbf{Input:} Initial wave function $\psi_0(\mathbf{r})$, 3D rotated atomic potentials $W$, slice separations $\{\Delta z_m\}_{m=1}^{N_z}$, defocus angles $\{\Delta f_j\}_{j=1}^{N_f}$, and interaction parameter $\sigma$.
    \begin{algorithmic}[1] 
      \State $\psi_{1}(\mathbf{r}) \gets \psi_0(\mathbf{r})$
      \For{$m\gets 1$ to $N_z$} \Comment{Beam propagation}
      \State $t_m(\mathbf{r})\gets \exp\left[i \sigma W_m(\mathbf{r})\right]$
      \State $g_m(\mathbf{r}) \gets t_m(\mathbf{r}) \cdot \psi_{m}(\mathbf{r})$
      \State $\psi_{m+1}(\mathbf{r}) \gets \mathcal{P}_{\Delta z_m}\left\{g_m(\mathbf{r})\right\}$
      \EndFor	
      \For{$j\gets 1$ to $N_f$} \Comment{Defocus and image}
      \State $\psi_{\rm{exit},j}(\mathbf{r}) \gets \mathcal{H}\left\{\mathcal{P}_{\Delta f_j}\left\{\psi_{m+1}(\mathbf{r})\right\}\right\}$      
      \EndFor	
    \end{algorithmic}
    \textbf{Return:} Predicted exit wave $\{\psi_{\rm{exit},j}(\mathbf{r})\}_{j=1}^{N_f}$ and intermediate wave function $\{\psi_m(\mathbf{r})\}_{m=1}^{N_z+1}$.
\end{algorithm}

To image the sample with minimal damage, a low dose is required. As a result, each individual measurement is very noisy. The measurement procedure can be abstracted to an electron counting process, with each measured pixel being modeled by Poisson noise with mean $\{\hat{I}_{i,j}(\mathbf{r})\}_{i=1,j=1}^{N_{\theta},N_f}$. Figures \ref{FigureSchematicSimulation}(f) and (g) illustrate a measurement process with a total electron budget of 7,000 electrons / \AA$^2$, which is equivalent to approximately 40 electrons / \AA$^2$ when distributed across 60 tilt angles having 3 defocused images each.

\subsection{Reconstruction Algorithm}
After we obtained all the intensity-only measurements, we estimate the potential $V$ by solving an optimization problem. For any estimated potential $V$, we can generate a series of predicted measurements $\{\hat{I}_{i,j}(\mathbf{r})\}_{i=1,j=1}^{N_{\theta},N_f}$ using the forward model. We formulate an error function to quantify the difference between predicted measurements the actual measurements $\{I_{i,j}(\mathbf{r})\}_{i=1,j=1}^{N_{\theta},N_f}$. Our goal is to find the optimal 3D atomic potential such that the error function is minimized:
\begin{equation} 
\label{eq:objective}
\begin{split}
	V &= \operatorname*{arg\,min}_{V}\displaystyle\sum_{i=1}^{N_{\theta}}\displaystyle\sum_{j=1}^{N_f}e_{i,j}^2 \\
    &= \operatorname*{arg\,min}_{V}\displaystyle\sum_{i=1}^{N_{\theta}}\displaystyle\sum_{j=1}^{N_f}\left\|\sqrt{I_{i,j}(\mathbf{r})} - \sqrt{\hat{I}_{i,j}(\mathbf{r})}\right\|^2_2,
\end{split}
\end{equation}
where $\|\cdot\|_2$ is the $l_2$ norm. Instead of directly comparing the difference between the predicted intensity measurements and true measurements, we compare the square roots of the intensity, which correspond to the amplitude of the exit waves. In a previous work~\cite{Yeh:15}, it was observed that the amplitude-based error function and the intensity-based error function produce the best reconstruction quality for Poisson-distributed noise and Gaussian-distributed noise in the measurements respectively. In this study, Poisson noise dominates when we use a low electron dose, and thus we have adopted an amplitude-based cost function.


We solve this optimization problem by applying an accelerated gradient method outlined in Algorithm \ref{alg:recon}. We first recursively compute the gradient for all tilt angles using back propagation, which is described in Algorithm \ref{alg:backward}. The mathematical derivation of Algorithm \ref{alg:backward}  is given in the appendix. Then, we perform a regularization process that enforces certain prior knowledge we have about the sample. The details of the regularization techniques are discussed later. Later on, we apply Nesterov's acceleration, which adds a momentum factor in the gradient update, to improve the convergence speed of our algorithm. 

Algorithms that assume lattice types and occupancies inevitably preclude detection of small scale spatial variation. Notice that during the reconstruction, we do not assume any structural priors of the sample. Thus, our method is robust enough to show vacancies and defects when they are present in the sample. In addition, different from \cite{VandenBroek2012method, VandenBroek:2013}, we do not assume specific shapes of the individual atoms.

\begin{algorithm}[H]
    \caption{Error backpropagation for gradient computation}
    \label{alg:backward}
    \textbf{Input:} Residual vectors $\{r_{j}(\mathbf{r})\}_{j=1}^{N_f}$, intermediate wave functions $\{\psi_{i,m}(\mathbf{r})\}_{m=1}^{N_z}$, 3D rotated atomic potentials $W$, slice separations $\{\Delta z_m\}_{m=1}^{N_z}$, defocus angles $\{\Delta f_j\}_{j=1}^{N_f}$, and interaction parameter $\sigma$.
    \begin{algorithmic}[1] 
     \State $\phi_{m+1}(\mathbf{r}) \gets 0$
      \For{$j\gets 1$ to $N_f$} \Comment{Refocus to end of sample}
      \State $\phi_{m+1}(\mathbf{r}) \gets \phi_{m+1}(\mathbf{r}) + \mathcal{P}_{-\Delta f_j}\left\{\mathcal{H}^{\dagger}\left\{r_{j}(\mathbf{r})\right\}\right\}$      
      \EndFor
      \For{$m\gets N_z$ to $1$} \Comment{Backpropagation}
      \State $t_m(\mathbf{r}) \gets \exp\left[-i \sigma W_m(\mathbf{r})\right]$
      \State $g_m(\mathbf{r}) \gets -i\sigma t_m(\mathbf{r}) \cdot \psi_{m}^*(\mathbf{r}) \cdot \phi_{m+1}(\mathbf{r})$
      \State $\phi_{m}(\mathbf{r}) \gets \mathcal{P}_{-\Delta z_m}\left\{t_m(\mathbf{r})\cdot\phi_{m+1}(\mathbf{r})\right\}$      
      \EndFor
    \end{algorithmic}
    \textbf{Return:} Estimated gradient $\nabla_V e_i \triangleq \{g_m(\mathbf{r})\}_{m=1}^{N_z}$.
\end{algorithm}

\begin{algorithm}[H]
    \caption{Iterative reconstruction}
    \label{alg:recon}
    \textbf{Input:} Tilt angles $\{\theta_i\}_{i=1}^{N_{\theta}}$, measured intensity images $\{I_{i,j}(\mathbf{r})\}_{i=1,j=1}^{N_{\theta},N_f}$, interaction parameter $\sigma$, stepsize $\alpha$, and maximum iteration $N_{it}$.
    \begin{algorithmic}[1] 
      \State $U^{(1)} \gets 0$, $V^{(0)} \gets 0$, $t^{(1)} = 1$
      \For{$k\gets 1$ to $N_{it}$} \Comment{Outer loop}
      \For{$i\gets 1$ to $N_{\theta}$} \Comment{Object rotation}
      \State $W_i = \mathcal{B}_{N_B}\left\{\mathcal{R}_{\theta_i}\left[U^{(k)}\right]\right\}$
      \State $\left(\{\psi_{\rm{exit},i,j}(\mathbf{r})\}_{j=1}^{N_f}, \{\psi_{i,m}(\mathbf{r})\}_{m=1}^{N_z+1}\right) \gets$ run Algorithm \ref{alg:forward} with $W_i$

\For{$j\gets 1$ to $N_f$} \Comment{Compute residual}
      \State $r_{i,j} \gets \psi_{\rm{exit},i,j} - \sqrt{I_{i,j}}\frac{\psi_{\rm{exit},i,j}}{\left|\psi_{\rm{exit},i,j}\right|}$
      \EndFor
	  \State $\nabla_Ve_i(U^{(k)}) \gets $ run Algorithm \ref{alg:backward} with $\{r_{i,j}(\mathbf{r})\}_{j=1}^{N_f}$, $\{\psi_{i,m}(\mathbf{r})\}_{m=1}^{N_z}$, and $W_i$
      \State $U^{(k)} \gets U^{(k)} - \alpha\mathcal{R}^{\dagger}_{\theta_i}\left\{\mathcal{B}_{N_B}^{\dagger}\left[\nabla_V e_i(U^{(k)})\right]\right\}$
      \EndFor
      \State $V^{(k)} \gets \text{prox}\left(U^{(k)}\right)$ \Comment{Regularization}
      \State $t^{(k+1)} \gets \frac{1+\sqrt{1+4(t^{(k)})^2}}{2}$ \Comment{Nesterov acceleration}
      \State $U^{(k+1)} \gets V^{(k)} + \left(\frac{t^{(k)}-1}{t^{(k+1)}}\right)(V^{(k)} - V^{(k-1)})$
      \EndFor
    \end{algorithmic}
    \textbf{Return:} Estimated atomic potential $V^{(k)}$.
\end{algorithm}

\subsection{Slice-Binning}
In both the forward and back propagation, the major bottleneck in computation is the Fourier transform. The number of Fourier transform performed is proportional to the number of slices in $z$. Since complete tomography without missing angles achieves isotropic resolution, the number of slices in $z$ should match the number of pixels reconstructed in $x$ and $y$, so the number of slices along the beam direction should be equally as dense, causing very heavy computation.

In this section, at every tilt angle we propose to increase the thickness of each slice (i.e. reducing axial resolution per angle). As a result, while total thickness of the sample remains constant, the total number of slices is reduced, along with the computation time. However, because tomography allows us to capture information about each voxel from multiple angles, the redundant information from the other tilt angles allows us to still reconstruct the object at atomic resolution isotropically.

In particular, we sum the 2D projected potentials of $N_B$ consecutive layers at each angle:
\begin{equation}
V_{B} = \mathcal{B}_{N_B}\{V\} = \left\{\sum_{m=1}^{N_B}V_{nN_B+m}(\mathbf{r})\right\}_{n=0}^{(N_z/N_B)-1}.
\end{equation}
We then compute both the forward model and back propagation using this binned potential. After the gradient is calculated, we distribute the gradient to the full volume by applying the adjoint operator, $\mathcal{B}^{\dagger}$:
\begin{equation}
V = \mathcal{B}_{N_B}^{\dagger}\{V_B\} = \left\{V_{B,\lceil \frac{m}{N_B} \rceil}(\mathbf{r})\right\}_{m=1}^{N_z},
\end{equation}
where $\lceil \cdot \rceil$ is the ceiling function. 

In the simulations shown in the results section, we bin every 10 slices. Since the pixel size in $z$ is 0.5 \AA, the effective slice separation becomes 5 \AA, which is sufficient to recover atomic resolution in the 2D parallel directions. This combined with many tilt angles will produce atomic resolution in 3D.

\subsection{Regularization}
Although the objective function in Equation \eqref{eq:objective} accounts for Poisson distribution of the electron counts in the measurements, the reconstruction can still be easily perturbed by noise. In addition, as we lower the amount of measurements, the inverse problem becomes ill-posed. Therefore, we propose a certain degree of regularization to alleviate this problem. In particular, we use the following regularized cost function:
\begin{equation}
\label{eq:objective_reg}
V = \operatorname*{arg\,min}_{V}\left\{\displaystyle\sum_{i=1}^{N_{\theta}}\displaystyle\sum_{j=1}^{N_f}e_{i,j}^2 + \alpha R(V)\right\},
\end{equation}
where $R(\cdot)$ is a general penalty function, and $\alpha$ is a tuning parameter to determine the strength of regularization. In this study, we have tested several common types of regularization methods. First, we tested Lasso (also known as $l_1$) regularization, where $R(V)=\left\|V\right\|_1$, that promotes sparsity in the natural domain, and is extensively used in statistical parameter estimations~\cite{Parikh:14}. We also investigated Total Variation (TV) regularization~\cite{Rudin:92}. Total Variation, where $R(V)=\left\|\mathcal{D}\{V\}\right\|_1$, where $\mathcal{D}\{\cdot\}$ denotes the finite difference operator, is a well-known denoising technique. Instead of enforcing sparsity in the natural domain, total variation enforces smoothness between neighboring pixels by promoting sparsity in the finite difference domain of the atomic potentials. This is a suitable candidate for regularization as we know \textit{a priori} that the 3D atomic potential is a smoothly varying function. We implement our regularization methods in a proximal gradient fashion. Outlined in Algorithm \ref{alg:recon}, we first compute the gradient sequentially using through-focus intensities captured from different angles. Then, we evaluate the proximal operator of the regularization techniques. Fortunately, Lasso regularization has an efficient closed-form evaluation. The evaluation for TV proximal operator, on the other hand, is in itself another iterative algorithm~\cite{Beck:09}.  In addition, since we know the atomic potential is purely real and positive (i.e. no absorption of electron beam), we can use the positivity constraint to refine our solution space. To enforce this constraint, we simply perform a projection of our estimate onto the real and positive space.

\subsection{Measurements of Atom Positions and Species}

After using the algorithm described above to reconstruct the atomic potentials, the final step is to measure the atomic coordinate positions and classify the atomic species based on the depth or size of the atomic potential well. We have adopted a similar atomic refinement strategy as previous AET studies~\cite{xu2015three, Yang:17}, which is referred to as ``atom tracing.'' First, the reconstructed volume is filtered with a smoothing kernel defined as a 3D Gaussian distribution with a standard deviation of 0.5 voxels minus another Gaussian distribution with a standard deviation of 1 voxel, normalized to zero total amplitude. Next, the local maxima are recorded as candidate atomic sites. These site positions are refined by fitting a 3D Gaussian function using nonlinear least squares. Next, the fitted intensities are subtracted from the reconstructed volume and candidate atomic sites are added by again filtering with a smoothing kernel and measuring local maximum.

Then an iterative fitting routine proceeds where for each atomic candidate, the nearest-neighbor site intensities are subtracted from the reconstructed volume. In this subtracted volume, nonlinear least squares is used to refine the 3D Gaussian function. After each of these iterations, several criteria were used to remove atomic coordinates.  Any sites with a very low intensity (below 30 V, approximately 10\% of the maximum sample potential) or size below ~1 voxel were removed, and any sites within 2.25 voxels of another site were merged into a single site. After approximately 12 refinement steps, each reconstruction trial was removing less than 2 atomic sites per iteration, and the root-mean square (RMS) change in atomic positions was less than 0.005 voxels. Note that the intensity and neighbor-distance thresholds we used were chosen to give good average performance across all datasets, and were not changed except in one specific instance described below. 

To classify atom species, we first generate a histogram of atom intensities. We then fit the histogram curve with a bi-modal Gaussian distribution, and we choose the intersection of the two Gaussian distributions to be the species classification threshold. All atoms having intensities less than the threshold will be classified as oxygen atoms, and the rest will be classified as silicon atoms. 

\begin{figure*}[!htbp]
	\centering
	\includegraphics[width=6.0in]{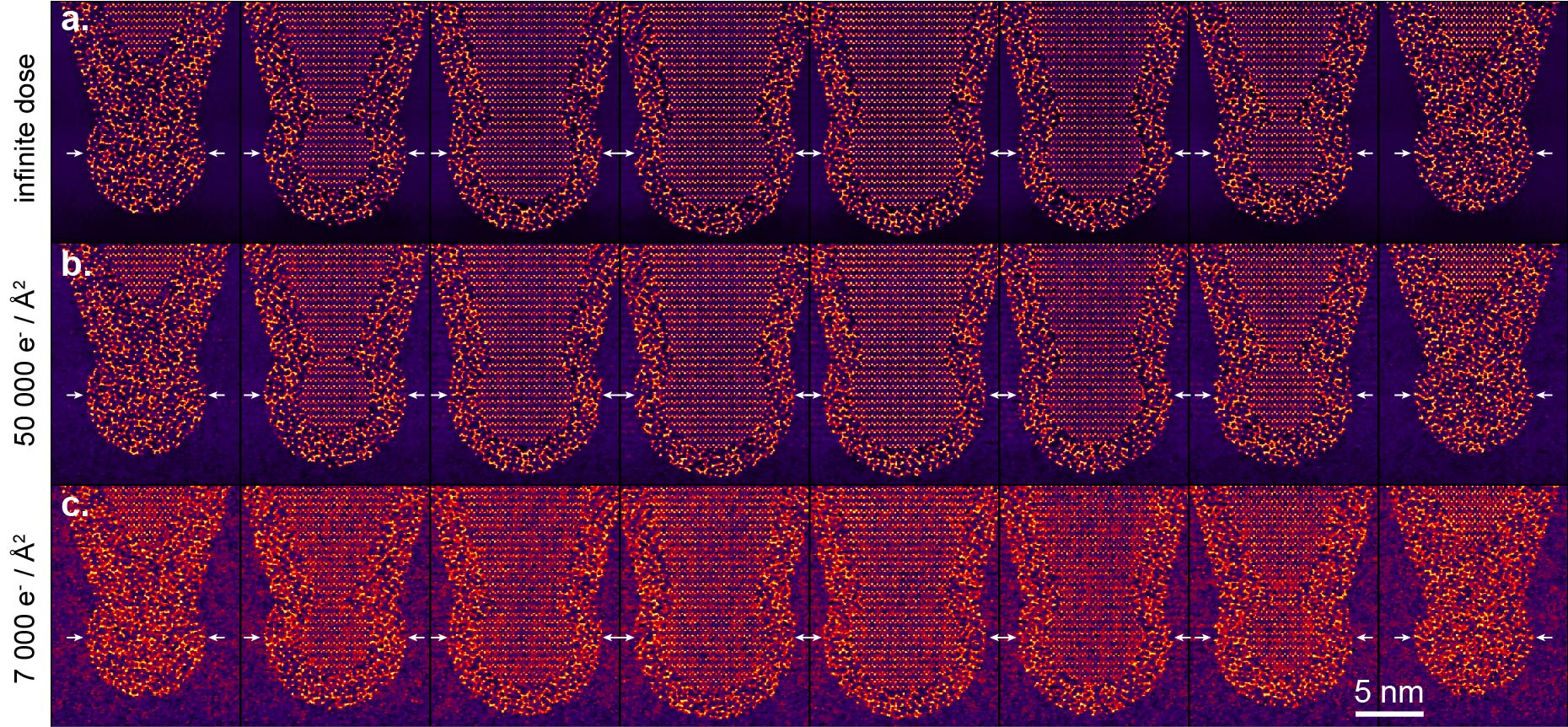}
	\caption{Two-dimensional 1 \AA\, thick slices of a simulated Si-SiO$_2$ reconstruction in  $x-y$ across multiple $z$ depth, using 36 tilt angles and 3 defocus values per tilt. (a) Infinite  dose, (b) 50 000 electrons / \AA$^2$, and (c) 7 000 electrons / \AA$^2$ total dose. Each slice shows the square root of the reconstructed potential from 0 to 80 volts, and tilt axis is along vertical direction. White arrows show location of reconstruction slices for the following sections.}
	\label{FigureRecon}
\end{figure*}

\subsection{Source Code and Results}

The forward simulation and reconstruction algorithms are implemented in Python, using Arrayfire package for GPU calculations. Atom tracing including position refinement and species determination, as well as visualizations were generated using Matlab codes. The atomic coordinates and reconstructed volumes are available online at [link to be added after publication].  The simulation and reconstruction codes are available at [link to be added after publication]. Atom tracing codes are available at [link to be added after publication].


\section{Results and Discussion}
\label{sec:results}

\begin{figure*}[!htbp]
	\centering
	\includegraphics[width=5.6in]{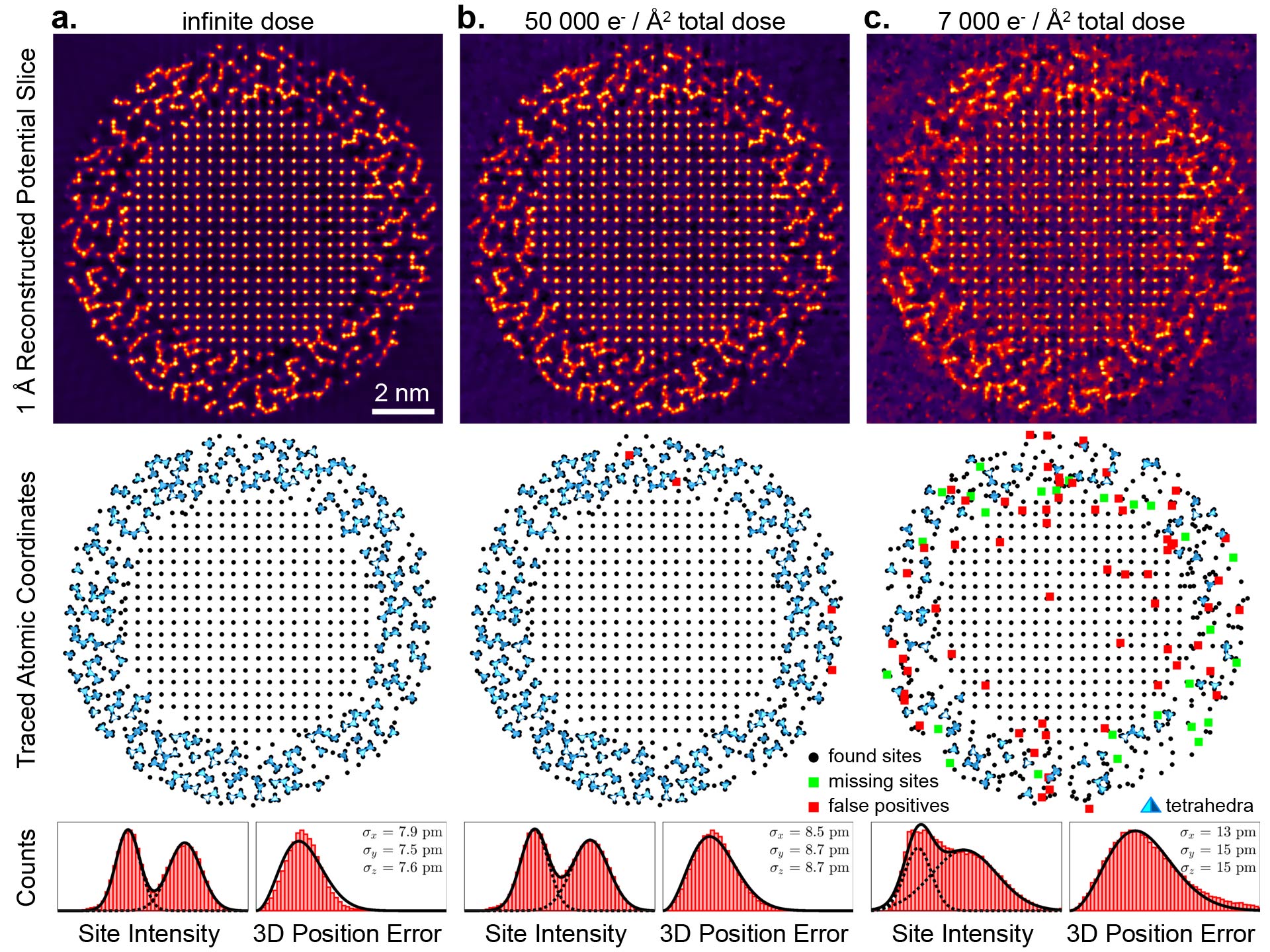}
	\caption{Phase contrast AET reconstructions for (a) infinite electron dose, (b) 50 000 electrons / $\mbox{\AA}^2$, and (c) 7 000 electrons / $\mbox{\AA}^2$ total dose.}
	\label{FigureResultsDose}
\end{figure*}

\begin{figure}[!htbp]
	\centering
		\includegraphics[width=3.3in]{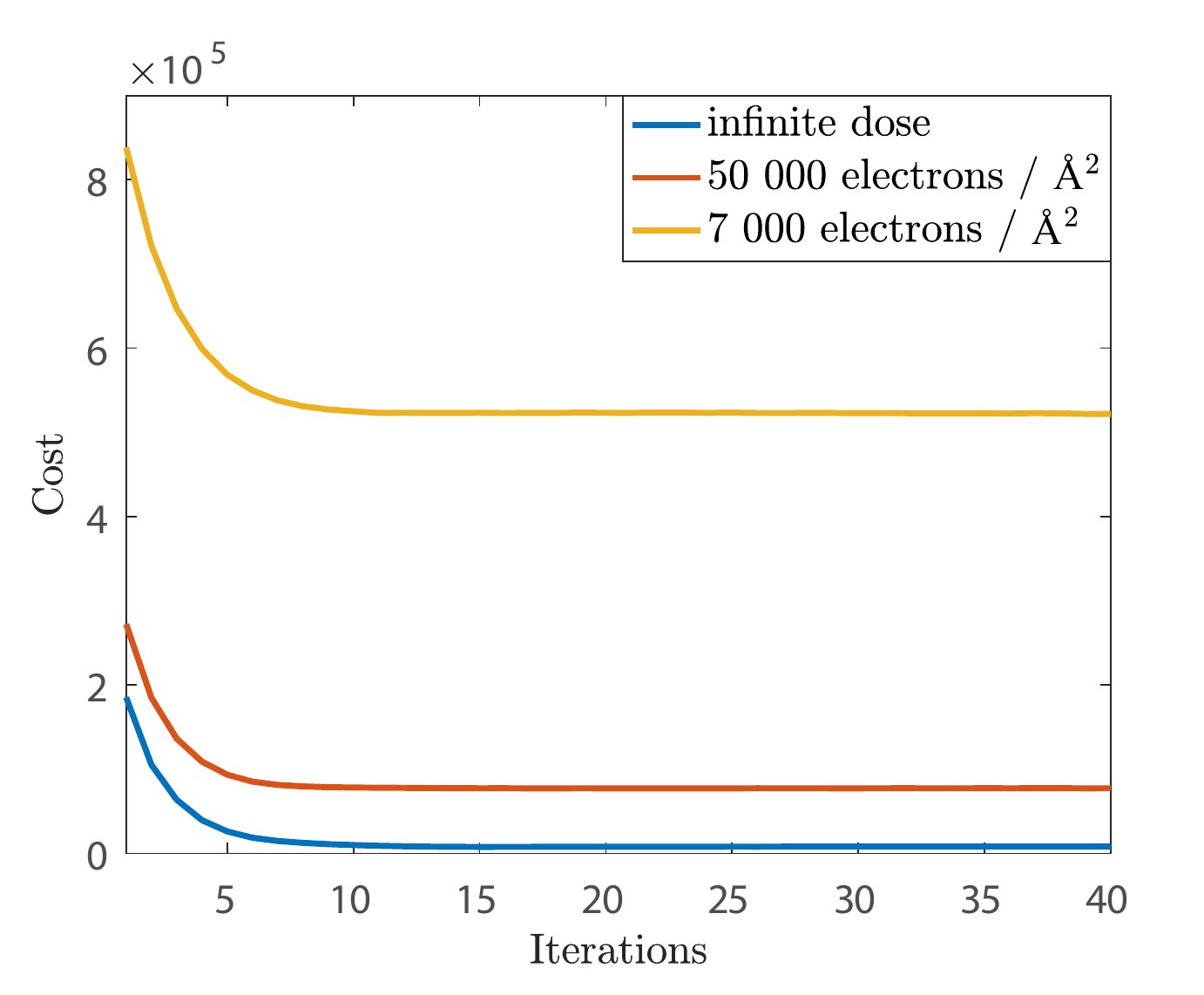}
	\caption{Plot of cost function vs iterations to show convergence for various dose budgets.}
	\label{FigureResultsCost}
\end{figure}

\begin{figure*}[!htbp]
	\centering
	\includegraphics[width=5.6in]{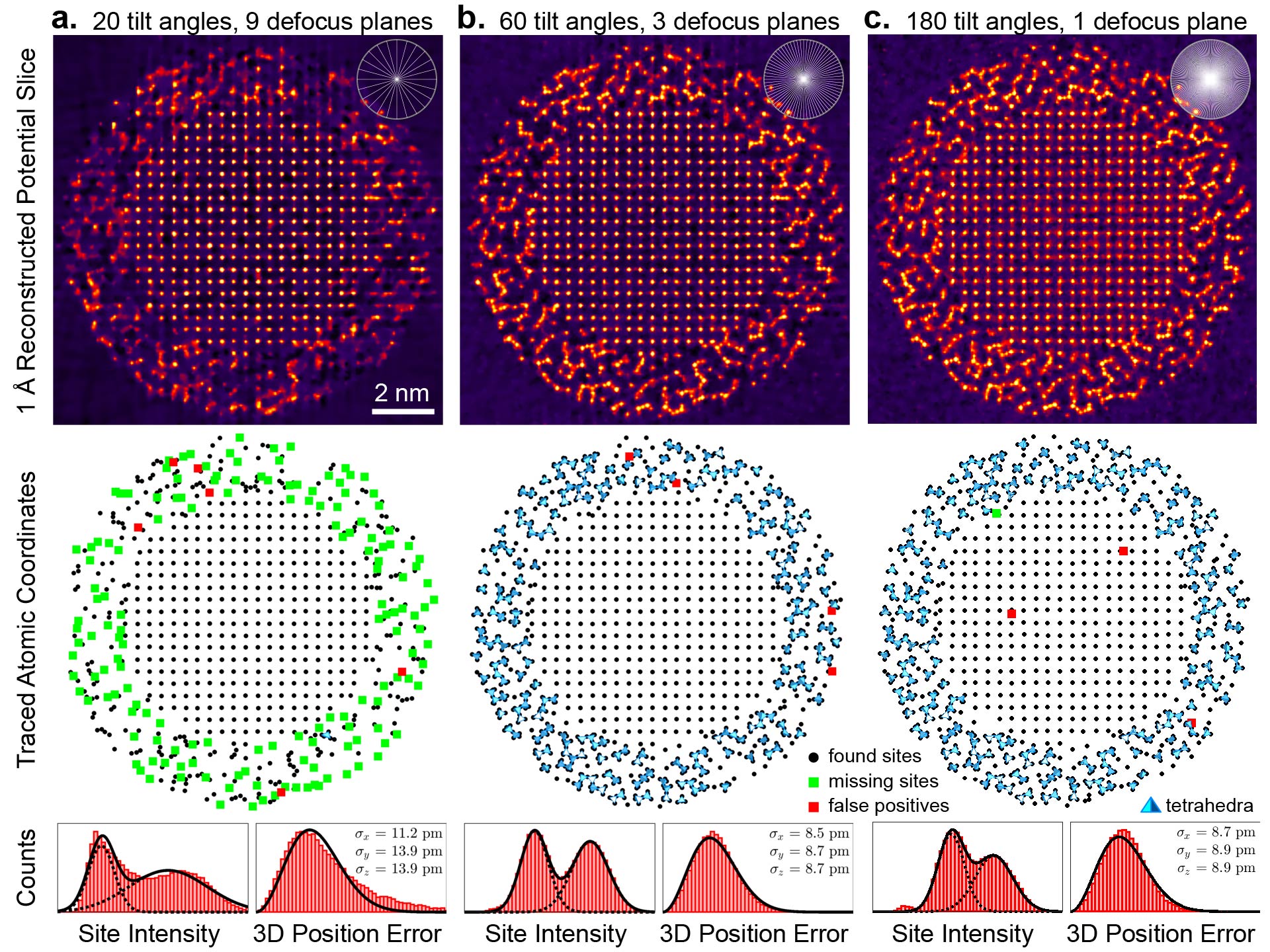}
	\caption{Phase contrast AET reconstructions for (a) 20 tilt angles with 9 defocus planes (linearly increasing from 20nm to 100nm), (b) 60 tilt angles with 3 defocus planes, and (c) 180 tilt angles with single defocus plane at 100nm.}
	\label{FigureResultsTiltDefocus}
\end{figure*}

\begin{figure*}[!htbp]
	\centering
	\includegraphics[width=5.6in]{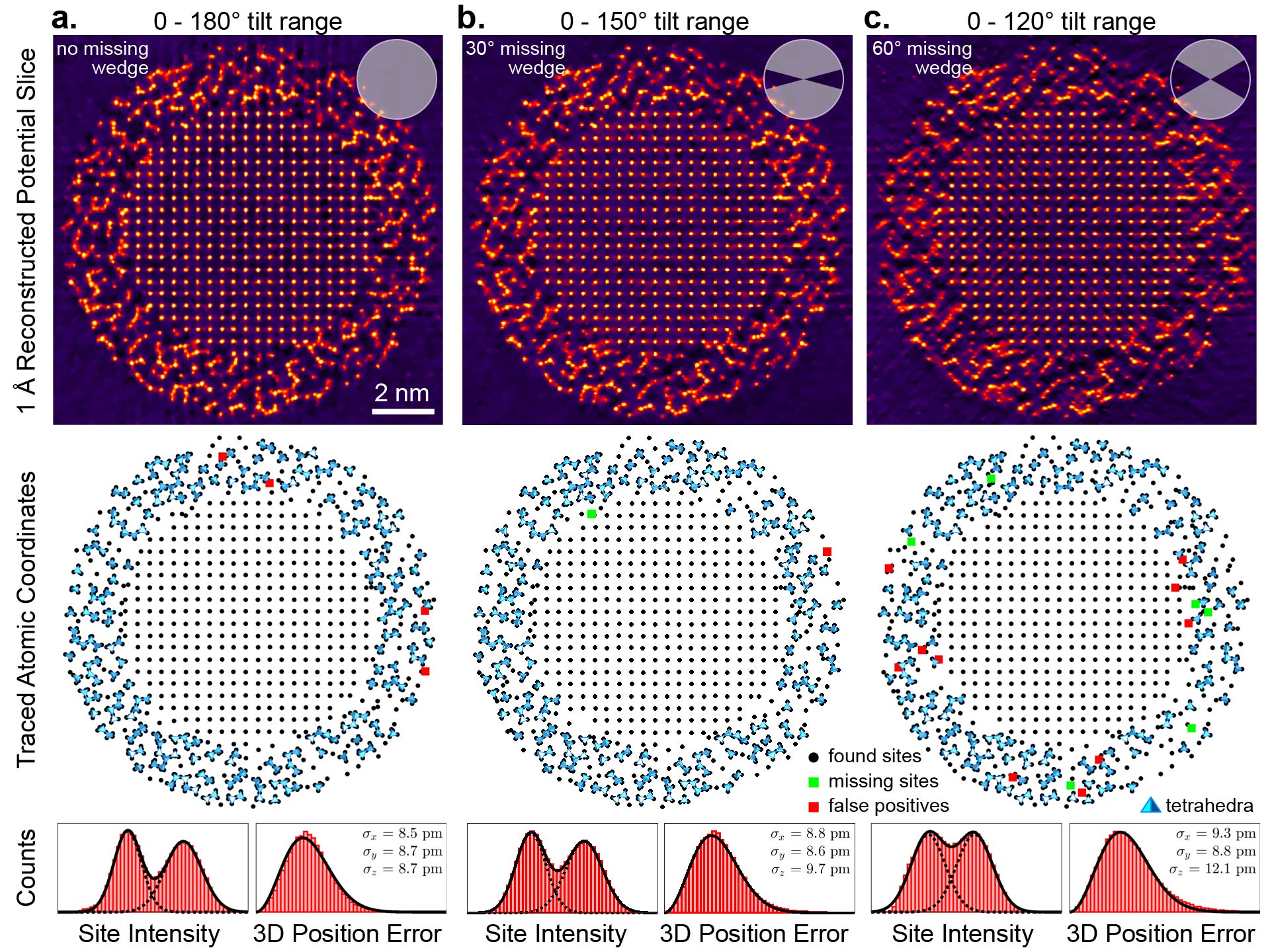}
	\caption{Phase contrast AET reconstructions for (a) full tomography data with no missing angle , (b) limited tomography data with 30$\degree$ missing angle, and (c) limited tomography data with 60$\degree$ missing angle.}
	\label{FigureResultsMissingWedge}
\end{figure*}

\begin{figure*}[!htbp]
	\centering
	\includegraphics[width=5.6in]{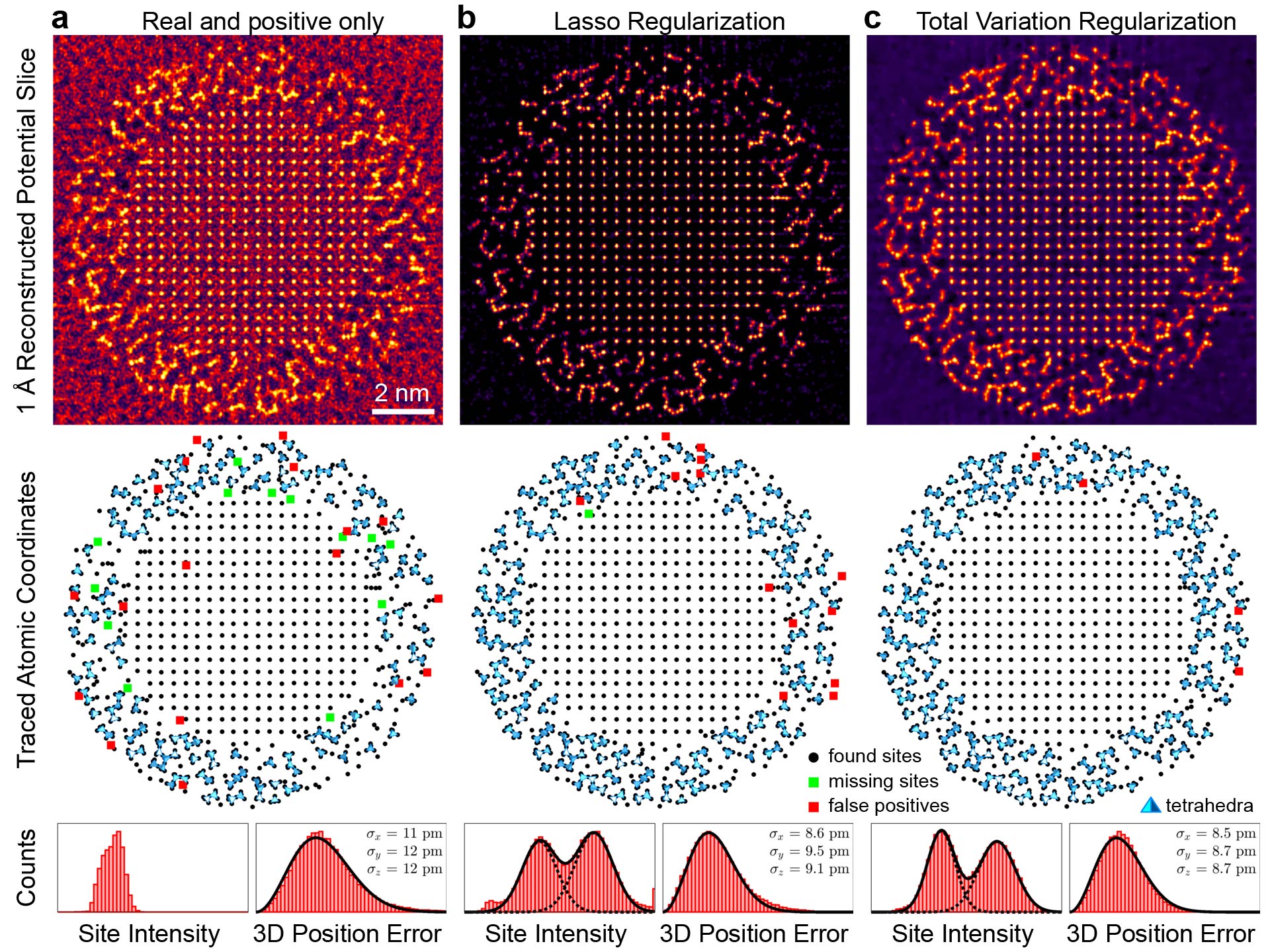}
	\caption{Phase contrast AET reconstructions using (a) real \& positivity constraints only, (b) Lasso regularization, and (c) total variation regularization.}
	\label{FigureResultsRegularization}
\end{figure*}
\begin{figure*}[!htbp]
	\centering
	\includegraphics[width=6.0in]{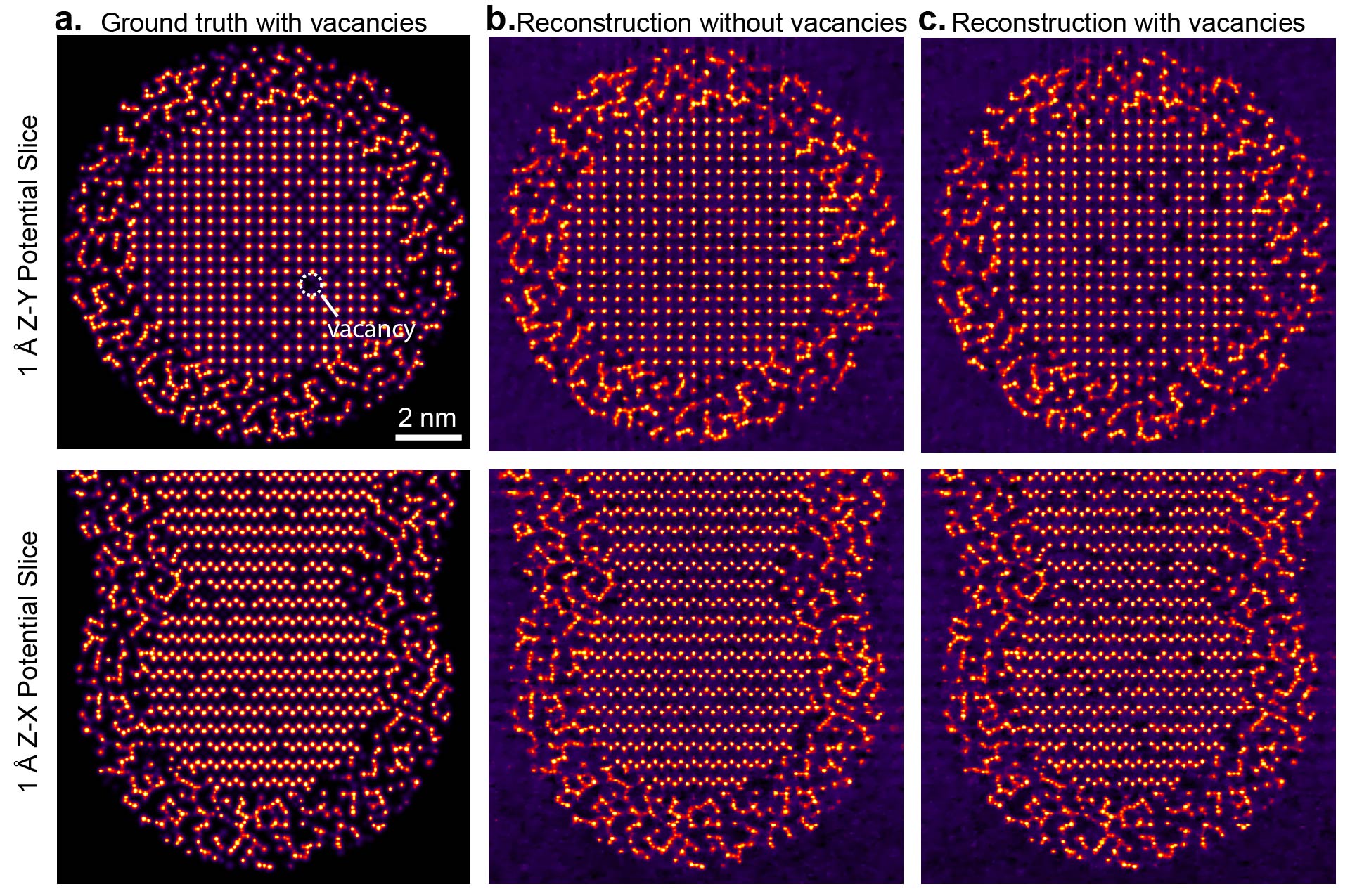}
	\caption{Phase contrast AET reconstructions when vacancies are introduced. (a) ground truth atomic potential with vacancies. Reconstruction when (b) no vacancies are present, and when (c) 30\% atoms are removed in the crystalline region and amorphous region. The top row shows the slices in $Z-Y$ direction, and the bottom row shows the slices in $Z-X$ direction.}
	\label{FigureResultsVacancy}
\end{figure*}

We have organized all results into several categories where we vary the experimental parameters. In each section, we test a single experimental or reconstruction parameter.

In each following set of reconstructions, a single slice of the reconstructed atomic potential that is perpendicular to the tilt axis is shown. The slice was taken from the thickest part of the protrusion, where the diameter is approximately 12 nm. For each slice, we have plotted the atomic coordinates that were correctly found, and the missing and false positives sites. Additionally, we have plotted tetrahedral shapes to each cluster of 5 atoms that formed a tetrahedron, with bond lengths of the 4 corner atoms to the center atom within 0.375 $\mbox{\AA}$ of the mean Si-O bond length of 1.6 $\mbox{\AA}$. These tetrahedra were added in order to help visualize how well the amorphous region of the sample was reconstructed, especially for reconstructions with a lot of noise or artifacts present. This feature classification is an example of the kind of classification measurement that could be performed even in the absence of clear atomic peaks, similar to that performed in structural biology \cite{kimanius2016accelerated}. Then, we show two histograms that quantify how well we trace the individual atoms. The first histogram shows the statistics of atomic potential intensities of identified atoms. The more resolved the two distributions are, the better we have classified the specific types of the atoms. The second histogram shows the errors of the 3D position estimation from the reconstruction. Here, for each identified atom we adopt the root-mean-square (RMS) from all coordinates:
\begin{equation}
\label{eq:positionerror}
\text{Position Error}_i = \sqrt{\left(x_i^*-\hat{x}_i\right)^2 + \left(y_i^*-\hat{y}_i\right)^2 + \left(z_i^*-\hat{z}_i\right)^2},
\end{equation}
where $x^*, y^*, z^*$ are the true coordinates, and $\hat{x}, \hat{y}, \hat{z}$ are the estimated coordinates. Ideally, a good reconstruction would show a histogram that has a peak close to 0, and a narrow main lobe. We also show RMS error ($\sigma$) in all three dimensions. 

All of the reconstructions in the following sections, unless otherwise stated, are full-angle TV regularized, created from 60 uniformly spaced tilt angles, each has 3 defocus intensity projections (25, 45, 100nm) with total incident electron count of 50 000 electrons / \AA$^2$. The regularization parameter $\alpha$ in \eqref{eq:objective_reg} is chosen such that the background noise is suppressed, without over-smoothing(smearing) the adjacent atoms. Additionally, for each of the scenarios mentioned, the algorithm is able to converge within 40 iterations. 

\subsection{Effect of Electron Dose}

In the first set of simulations, reconstructions using three different dose budgets are compared to examine how noise affects the algorithm performance. In particular, infinite electron dose (noiseless), 50 000 electrons / \AA$^2$, and 7 000 electrons / \AA$^2$ were chosen. Figure \ref{FigureRecon}(a)-(c) show lateral slices at multiple $z$ depths of the reconstructed volume taken from simulations with different dose level. Figure \ref{FigureResultsDose} shows a single $z$-$x$ cross-section slice, where the location is indicated by the white arrows in Figure \ref{FigureRecon}, and atom tracing results. In all three reconstructions, the atomic peaks can be easily identified. The reconstruction using 50 000 electrons / \AA$^2$ total dose over all tilts and defocused images is essentially identical to the infinite dose reconstruction. 

However, as expected, the reconstruction quality deteriorates as we decrease the dose budget, with the background becoming noticeably more noisy. In the meantime, we cannot further increase the regularization as it will over-smooth the reconstruction. For the dose level of 7000 electrons / \AA$^2$, atoms that are too close to each other are smeared together, and so missing sites increase substantially. Noisy fluctuations in the background lead to an increased number of false positive sites. The noise perturbation also causes loss of contrast in the atomic potential intensity. This loss can be seen from the intensity histogram: the intensity distributions of two types of atoms are less resolved when dose is decreased, making it harder to classify the species of individual atoms. Finally, the RMS error for position estimation increases isotropically when we decrease the dose level. 

Figure \ref{FigureResultsCost} shows the plot of cost function \eqref{eq:objective} vs iterations. Despite the convergence, as we lower the dose budget, the predicted intensity of the reconstruction has more mismatch with the measured intensity, causing the squared error to increase. Due to lack of further insights, we do not show convergence plots for the rest of the set of reconstructions.

\subsection{Effect of Number of Tilt and Defocus Measurements}

Because total dose is distributed across measurements from all tilt angles and defocus distances, we face the trade off between number of tilt angles ($N_\theta$) and defocus planes ($N_f$). Therefore, in this set of simulations, we compare the performance difference of our proposed method as we vary $N_\theta$ and $N_f$, while keeping the dose level the same, equal to 50 000 e / \AA$^2$. For the first configuration, as shown in Figure \ref{FigureResultsTiltDefocus}(a)-(c) are reconstructions from 20 uniformly spaced projections (separated by 1$\degree$) with 9 defocus planes (20 nm-100 nm in steps of 10 nm), 60 projections with 3 defocus planes (20 nm, 45 nm, and 100 nm), and 180 projections with a single plane at 100 nm, respectively. These defocus values were chosen by using numerical testing to find a good balance between using larger defocus values to produce more contrast, but not large enough to make image alignment difficult or lose resolution due to coherence limits.

Clearly, by trading the number of defocus planes for more tilt angles, we are able to better reconstruct the sample's structure, and perform improved atom tracing. Comparing Figure \ref{FigureResultsTiltDefocus}(a) and (b), we notice that including more tilt angles will generate less missing sites, especially in the amorphous SiO$_2$ region. However, we also notice the diminishing returns as we increase the number of tilts. On the other hand, the advantage of through-focal measurements at each tilt angle is that we can better quantitatively reconstruct the atomic potential of the 3D object \cite{Zhong:14}. Comparing the site intensity histograms in Figure \ref{FigureResultsTiltDefocus}(b) and (c), we see that the distributions of the silicon and oxygen atoms are more resolved, so we can achieve atom classification with a lower error rate.

\subsection{Effect of Missing Tilt Angles}

When full-angle tomography dataset is available, isotropic resolution can be achieved in $x$, $y$, and $z$. However, when projection angles are missing due to sample geometry or sample stage, the coverage of object's Fourier spectrum is incomplete \cite{muller2015theory}, and the uncovered region is described by a ``missing wedge". Next, we test our algorithm on 30$\degree$ and 60$\degree$ missing wedge tilt series, giving the reconstructions shown in Figures \ref{FigureResultsMissingWedge}(b) and (c) respectively. In the simulation setup, the angle separation is constant. Therefore, when the same total dose (again 50 000 e / \AA$^2$) is distributed across all acquisitions, the dose per image increases.  

As we increase the amount of missing angles, the axial resolution deteriorates along the missing wedge direction. This reconstruction degradation increases the errors in atom tracing and identification.  Comparing the reconstruction in \ref{FigureResultsMissingWedge}(a) to that in (c), the portion of missing sites increases from 0.06\% to 0.98\%. Not only is it harder to identify atoms, it is also more challenging to correctly identify the 3D positions of each atom. The position error histogram in Figure \ref{FigureResultsMissingWedge}(c) suggests that position estimate is very inaccurate in the axial direction as we increase the missing wedge, while the estimation accuracy in the lateral directions are maintained.

\subsection{Effect of Regularization}

Because low dose is required in order to preserve sample structure during imaging, Poisson noise dominates and severely perturbs the reconstruction. Therefore, in the last set of simulation we examine the effectiveness that different regularization techniques have on the reconstruction. In particular, we compare pure positivity \& real constraint, Lasso regularization, and total variation regularization, which are shown in Figure \ref{FigureResultsRegularization}(a)-(c) respectively.

The results suggest that some degree of regularization is necessary for low dose measurements. With only real \& positivity constraint, the background is too noisy to perform accurate atom tracing. The intensity histogram shows that it fails to provide two resolved peaks that are needed to perform atom classification. The position estimation error in all directions is larger compared to the other two methods. 

Both Lasso and TV regularization produce a high quality reconstruction. The Lasso reconstruction produces sharper peaks, but also tends to shrink some peak intensities as well as sizes of the potential wells. This leads to a worse distribution of peak intensities, making atomic species classification less accurate. The peak positions are also approximately 10\% less accurate. Therefore we have selected TV regularization as the standard for our reconstructions. 

\subsection{Vacancies in crystalline Si and amorphous SiO$_2$}

The algorithm that we propose is capable of identifying single-atom defects or vacancies in the sample. Here, we validate this claim by simulating the Si-SiO$_2$ tip sample that contains vacancies. We simulate the vacancies and defects by randomly removing approximately $5\%$ of the atoms in the original sample. Then, with the same geometry and experimental configuration as that in Figure \ref{FigureResultsDose}(b), we reconstruct the atomic potentials of the defected sample. Figure \ref{FigureResultsVacancy}(a) shows the ground truth atomic potential after the atoms have been removed. The reconstruction result is shown in Figure \ref{FigureResultsVacancy}(c). We also refer to Figure \ref{FigureResultsVacancy}(b) for the case where no atoms are removed. Samples can still be reconstructed when there are single-atom defects present, because the algorithm does not assume any structural priors. 

\subsection{Summary of Reconstruction Results}

\begin{table*}[ht]
\caption{Summary of atom tracing results, out of 62 402 sites in the tip region with a radius $\leq$ 12 nm diameter.}
\centering
\begin{tabular}{m{0.10\linewidth}m{0.14\linewidth}m{0.08\linewidth}m{0.09\linewidth}m{0.08\linewidth}m{0.08\linewidth} | m{0.08\linewidth}m{0.08\linewidth}m{0.08\linewidth}m{0.08\linewidth}}
\hline
Figure(s) 
& Total Dose 
& \# Tilts 
& \# Defocus 
& Tilt Span
& Reg. 
& Position Error
& Atoms Found
& False Positives 
& Correct Species 
\\ \hline
	\ref{FigureResultsDose}(a) &  
    infinite & 60 & 3 & 180$\degree$ & TV & 
    \multicolumn{1}{r}{12.51 pm} & 
    \multicolumn{1}{r}{99.98\%} & 
    \multicolumn{1}{r}{0.035\%} & 
    \multicolumn{1}{r}{98.63\%} 
\\
    {\bf  \ref{FigureResultsDose}(b) } 
    & 
    50 000 e / \AA$^2$ & 60 & 3 & 180$\degree$ & TV & 
    \multicolumn{1}{r}{13.91 pm} & 
    \multicolumn{1}{r}{99.94\%} & 
    \multicolumn{1}{r}{0.181\%} & 
    \multicolumn{1}{r}{96.44\%}
\\
    \ref{FigureResultsDose}(c) 
    & 
    7 000 e / \AA$^2$ & 60 & 3 & 180$\degree$ & TV & 
    \multicolumn{1}{r}{21.62 pm} & 
    \multicolumn{1}{r}{95.24\%} & 
    \multicolumn{1}{r}{9.583\%} & 
    \multicolumn{1}{r}{79.79\%} 
\\ \hline
    \ref{FigureResultsTiltDefocus}(a) 
    & 
    50 000 e / \AA$^2$ & 20 & 9 & 180$\degree$ & TV & 
    \multicolumn{1}{r}{19.11 pm} & 
    \multicolumn{1}{r}{72.48\%} & 
    \multicolumn{1}{r}{1.191\%} & 
    \multicolumn{1}{r}{82.71\%} 
\\
    {\bf \ref{FigureResultsTiltDefocus}(b) } 
    & 
    50 000 e / \AA$^2$ & 60 & 3 & 180$\degree$ & TV & 
    \multicolumn{1}{r}{13.91 pm} & 
    \multicolumn{1}{r}{99.94\%} & 
    \multicolumn{1}{r}{0.181\%} & 
    \multicolumn{1}{r}{96.44\%}
\\
	\ref{FigureResultsTiltDefocus}(c) 
    & 
    50 000 e / \AA & 180 & 1 & 180$\degree$ & TV & 
    \multicolumn{1}{r}{14.30 pm} & 
    \multicolumn{1}{r}{99.97\%} & 
    \multicolumn{1}{r}{0.812\%} & 
    \multicolumn{1}{r}{91.15\%} 
\\ \hline
    {\bf \ref{FigureResultsMissingWedge}(a) } 
    & 
    50 000 e / \AA$^2$ & 60 & 3 & 180$\degree$ & TV & 
    \multicolumn{1}{r}{13.91 pm} & 
    \multicolumn{1}{r}{99.94\%} & 
    \multicolumn{1}{r}{0.181\%} & 
    \multicolumn{1}{r}{96.44\%}
\\
	\ref{FigureResultsMissingWedge}(b) 
    & 
    50 000 e / \AA$^2$ & 60 & 3 & 150$\degree$ & TV & 
    \multicolumn{1}{r}{14.34 pm} & 
    \multicolumn{1}{r}{99.75\%} & 
    \multicolumn{1}{r}{0.450\%} & 
    \multicolumn{1}{r}{94.67\%} 
\\
	\ref{FigureResultsMissingWedge}(c) 
    & 
    50 000 e / \AA & 60 & 3 & 120$\degree$ & TV & 
    \multicolumn{1}{r}{15.84 pm} & 
    \multicolumn{1}{r}{99.02\%} & 
    \multicolumn{1}{r}{1.960\%} & 
    \multicolumn{1}{r}{90.50\%} 
\\ \hline
	\ref{FigureResultsRegularization}(a) 
    & 
    50 000 e / \AA$^2$ & 60 & 3 & 180$\degree$ & positive & 
    \multicolumn{1}{r}{18.65 pm} & 
    \multicolumn{1}{r}{97.81\%} & 
    \multicolumn{1}{r}{1.647\%} & 
    \multicolumn{1}{r}{46.81\%} 
\\
	\ref{FigureResultsRegularization}(b) 
    & 
    50 000 e / \AA$^2$ & 60 & 3 & 180$\degree$ & Lasso & 
    \multicolumn{1}{r}{14.17 pm} & 
    \multicolumn{1}{r}{99.78\%} & 
    \multicolumn{1}{r}{0.729\%} & 
    \multicolumn{1}{r}{92.53\%} 
\\
    {\bf \ref{FigureResultsRegularization}(c) } 
    & 
    50 000 e / \AA$^2$ & 60 & 3 & 180$\degree$ & TV & 
    \multicolumn{1}{r}{13.91 pm} & 
    \multicolumn{1}{r}{99.94\%} & 
    \multicolumn{1}{r}{0.181\%} & 
    \multicolumn{1}{r}{96.44\%}
\\
\hline
\end{tabular}
\label{table:results}
\end{table*}

Finally, in Table \ref{table:results}, we list all atom tracing and classification results of the earlier reconstructions. Notice that cases in Figure \ref{FigureResultsDose}(b), Figure \ref{FigureResultsTiltDefocus}(b), \ref{FigureResultsMissingWedge}(a), and Figure \ref{FigureResultsRegularization}(c) are equivalent and are highlighted. We report mean 3D position error (smaller is better), portion of the atoms correctly found (larger is better), portion of false positives (smaller is better), and the portion of atoms where the species are correctly labeled (larger is better).

\section{Conclusion}
\label{sec:conclusion}

In this paper, we have described a reconstruction algorithm for atomic electron tomography, from a tilt series of defocused plane-wave HRTEM images. Our nonlinear model, with schematics shown in Figure \ref{FigureSchematicGeometry}(a) takes into account the multiple scattering events of the electron beam. We use slice-binning and fast rotation and propagation algorithms to decrease the reconstruction time, and TV regularization to improve the reconstruction quality.  Using a sample with both crystalline Si and amorphous SiO$_2$ in a core-shell tip geometry, we have demonstrated accurate atomic reconstructions of more than 60 000 atoms in a sample with a diameter up to 12 nm. We also showed that our method is robust to low dose measurements, works for a small number of defocused images or even a single image per tilt angle, and can handle a missing wedge in the tilt angles of up to 60$\degree$. Our method will enable atomic-resolution tomographic reconstruction of nanoscale samples for samples containing both strongly and weakly-scattering elements, with either crystalline or amorphous structures. Finally, we are releasing all source codes to encourage researchers to perform phase contrast atomic electron tomography experiments to solve structures on the smallest length scales.

\section{Acknowledgements}
We thank Jihan Zhou, Mary Scott, and Hannah DeVyldere for helpful discussions. This work was supported by STROBE: A National Science Foundation Science \& Technology Center under Grant No. DMR 1548924. Work at the Molecular Foundry was supported by the Office of Science, Office of Basic Energy Sciences, of the U.S. Department of Energy under Contract No. DE-AC02-05CH11231. We also wish to thank the NVIDIA Corporation for donation of GPU resources.
\appendix
\section{Appendix}
\subsection{Gradient derivation}
\label{app:gradient}
In this section, we derive the details of our approach to solve for the inverse problem in vectorized notation. First, we discretize the coordinate system into $N_x$ and $N_y$ pixels for $\mathbf{r} = (x,y)$ respectively. We sample all 2D functions at these discrete coordinates. Then, we raster-scanned the samples into column vectors in $\mathbb{R}^{N_xN_y}$. In addition, linear operators $\mathcal{H},\mathcal{P},\mathcal{F}$ can be represented by matrices $\mathbf{H},\mathbf{P},\mathbf{F} \in \mathbb{C}^{N_xN_y \times N_xN_y}$

For a given tilt angle $\theta_i$ and defocus $\Delta f_j$ measurement, the error function in \eqref{eq:objective} can be expressed as:
\begin{eqnarray}
e_{i,j}^2  = \mathbf{e}_{i,j}^{\dagger}\mathbf{e}_{i,j}
\end{eqnarray}
where $\mathbf{e}_{i,j} = \sqrt{\mathbf{I}_{i,j}} - \sqrt{\mathbf{\hat{I}}_{i,j}}$, and $(\cdot)^{\dagger}$ is the hermitian adjoint of a matrix or a vector. 

Because the multislice propagation model assumes that the atomic potentials of each layer is independent of each other, we calculate the derivative of $e_{i,j}^2$ with respect to every layer of the potentials $\mathbf{V}_m$ separately  by applying chain rule:
\begin{equation}
\begin{split}
\nabla_{\mathbf{V}_m}e_{i,j}^2(\mathbf{V}_m) & =
\left[\frac{\partial\mathbf{e}_{i,j}^{\dagger}\mathbf{e}_{i,j}}{\partial\mathbf{V}_m}\right]^{\dagger}  =  \left[\frac{\partial\mathbf{e}_{i,j}^{\dagger}\mathbf{e}_{i,j}}{\partial\mathbf{e}_{i,j}}\frac{\partial\mathbf{e}_{i,j}}{\partial\mathbf{V}_m}\right]^{\dagger} \\
& = 
\left[2\mathbf{e}_{i,j}\frac{\partial\mathbf{e}_{i,j}}{\partial\mathbf{V}_m}\right]^{\dagger}.
\end{split}
\end{equation}
Next, we show the calculation of $\frac{\partial\mathbf{e}_{i,j}}{\partial\mathbf{V}_m}$ using backpropagation. Following \eqref{eq:slice_prop}  and \eqref{eq:exitwave}, the derivative of $\mathbf{e}_{i,j}$ with respect to the $m^{th}$ layer $\mathbf{V}_m$ is:
\begin{equation}
\begin{split}
\frac{\partial\mathbf{e}_{i,j}}{\partial\mathbf{V}_m} 
= -&\frac{\partial(|\mathbf{\psi}_{\rm{exit},j}|^2)^{1/2}}{\partial|\mathbf{\psi}_{\rm{exit},j}|^2}
\frac{\partial\text{diag}(\mathbf{\psi}_{\rm{exit},j}^*)\mathbf{\psi}_{\rm{exit},j}}{\partial\mathbf{\psi}_{N_z+1}}\\
& \frac{\partial\mathbf{\psi}_{N_z+1}}{\partial\mathbf{\psi}_{N_z}}
\cdots\frac{\partial\mathbf{\psi}_{m+1}}{\partial\mathbf{t}_{m}}
\frac{\partial\mathbf{t}_{m}}{\partial\mathbf{V}_{m}},
\end{split}
\end{equation}
where $(\cdot)^*$ denotes complex conjugate, $\text{diag}(\cdot)$ is an operator that puts a vector into the diagonal of a square matrix, 
\begin{eqnarray}
\frac{\partial(|\mathbf{\psi}_{\rm{exit},j}|^2)^{1/2}}{\partial(|\mathbf{\psi}_{\rm{exit},j}|^2)} & = & \frac{1}{2}\text{diag}\left(\frac{1}{|\mathbf{\psi}_{\rm{exit},j}|}\right),\\
\frac{\partial\text{diag}(\mathbf{\psi}_{\rm{exit},j}^*)\mathbf{\psi}_{\rm{exit},j}}{\partial\mathbf{\psi}_{N_z+1}} & = & \text{diag}(\mathbf{\psi}_{\rm{exit},j}^*)\mathbf{H}\mathbf{P}_{\Delta f_j},\\
\frac{\partial\mathbf{\psi}_{N_z+1}}{\partial\mathbf{\psi}_{N_z}} & = & \mathbf{P}_{\Delta z_{N_z}} \text{diag}(\mathbf{t}_{N_z}),\\ 
\frac{\partial\mathbf{\psi}_{m+1}}{\partial\mathbf{t}_{m}} & = & \mathbf{P}_{\Delta z_m} \text{diag}(\mathbf{\psi}_{N_z}), \mathrm{and} \\
\frac{\partial\mathbf{t}_{m}}{\partial\mathbf{V}_{m}} & = & i\sigma\text{diag}(\mathbf{t}_m).
\end{eqnarray}
Combining the terms, we arrive at the gradient of $e_{i,j}^2$ with respect to $\mathbf{V}_m$:
\begin{equation}
\begin{split}
\nabla_{\mathbf{V}_m}&e_{i,j}^2(\mathbf{V}_m)=\\
&i\sigma\text{diag}(\mathbf{t}_m^*\cdot\mathbf{\psi}_{N_z}^*)\mathbf{P}_{-\Delta z_m}\cdots \text{diag}(\mathbf{t}_{N_z}^*)\mathbf{P}_{-\Delta z_{N_z}} \\ 
&\mathbf{P}_{-\Delta f_j}\mathbf{H}^{\dagger}\text{diag}\left(\frac{\mathbf{\psi}_{\rm{exit},j}}{|\mathbf{\psi}_{\rm{exit},j}|}\right)\left(\sqrt{\mathbf{I}_{i,j}} - \sqrt{\mathbf{\hat{I}}_{i,j}}\right)
\end{split}
\end{equation}
If we consider all defocus measurements at tilt angle $\theta_i$, the gradient then becomes:
\begin{equation}
\begin{split}
\nabla_{\mathbf{V}_m}&e_{i,j}^2(\mathbf{V}_m)=\\
&i\sigma\text{diag}(\mathbf{t}_m^*\cdot\mathbf{\psi}_{N_z}^*)\mathbf{P}_{-\Delta z_m}\cdots \text{diag}(\mathbf{t}_{N_z}^*)\mathbf{P}_{-\Delta z_{N_z}} \\ 
&\displaystyle\sum_{j=1}^{N_f}\mathbf{P}_{-\Delta f_j}\mathbf{H}^{\dagger}\text{diag}\left(\frac{\mathbf{\psi}_{\rm{exit},j}}{|\mathbf{\psi}_{\rm{exit},j}|}\right)\left(\sqrt{\mathbf{I}_{i,j}} - \sqrt{\mathbf{\hat{I}}_{i,j}}\right).
\end{split}
\end{equation}
The specific steps for computing the gradient is described in Algorithm \ref{alg:backward} and Algorithm \ref{alg:recon}.







\subsection{Slice-Binning Validation}
\label{app:slicebinning}

\begin{figure}[!htbp]
	\centering
		\includegraphics[width=3.3in]{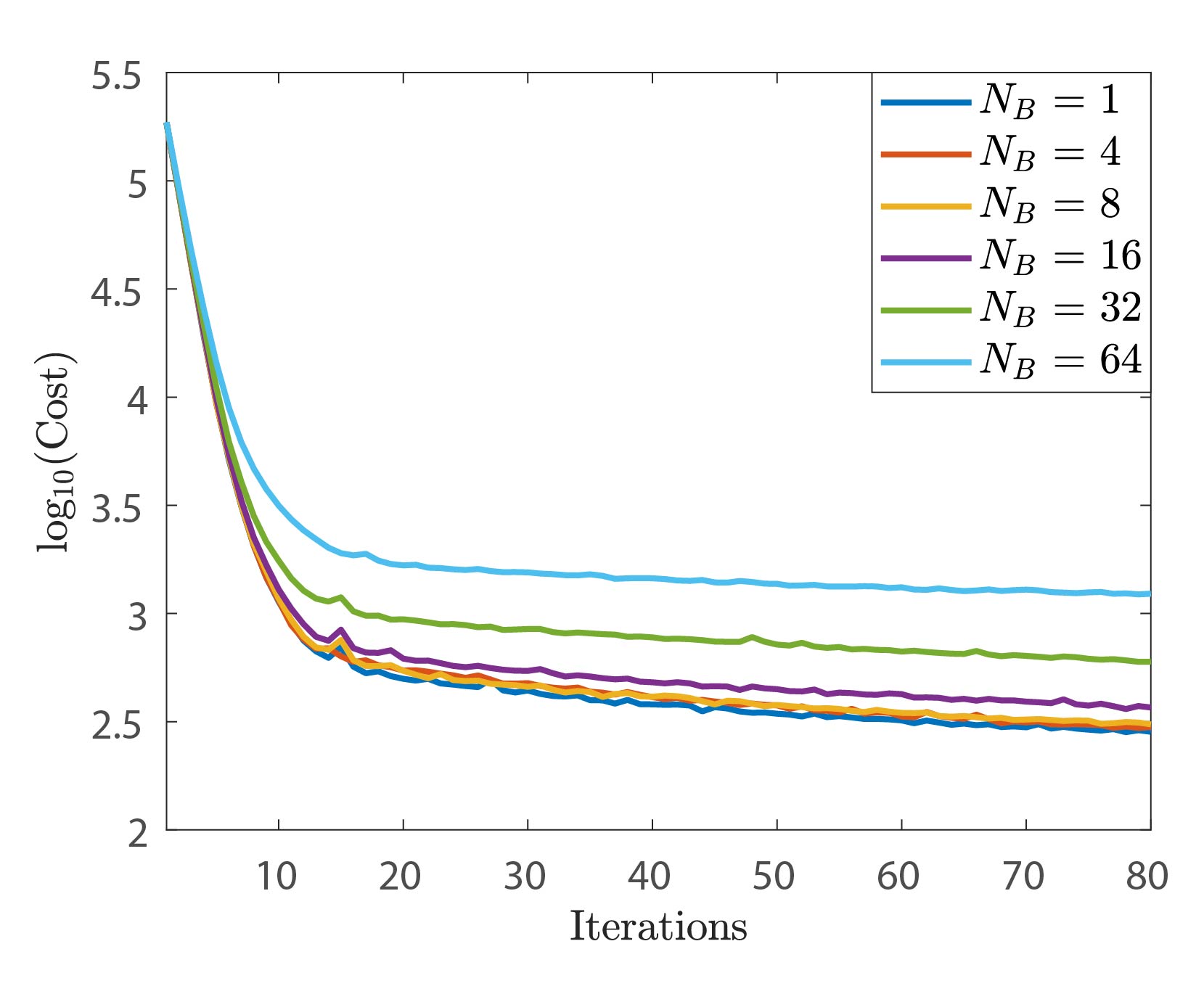}
	\caption{Plot of cost function vs iterations to show convergence for various binning factors ($N_B$).}
	\label{FigureResultsBinning_Convergence}
\end{figure}

\begin{figure}[!htbp]
	\centering
		\includegraphics[width=3.3in]{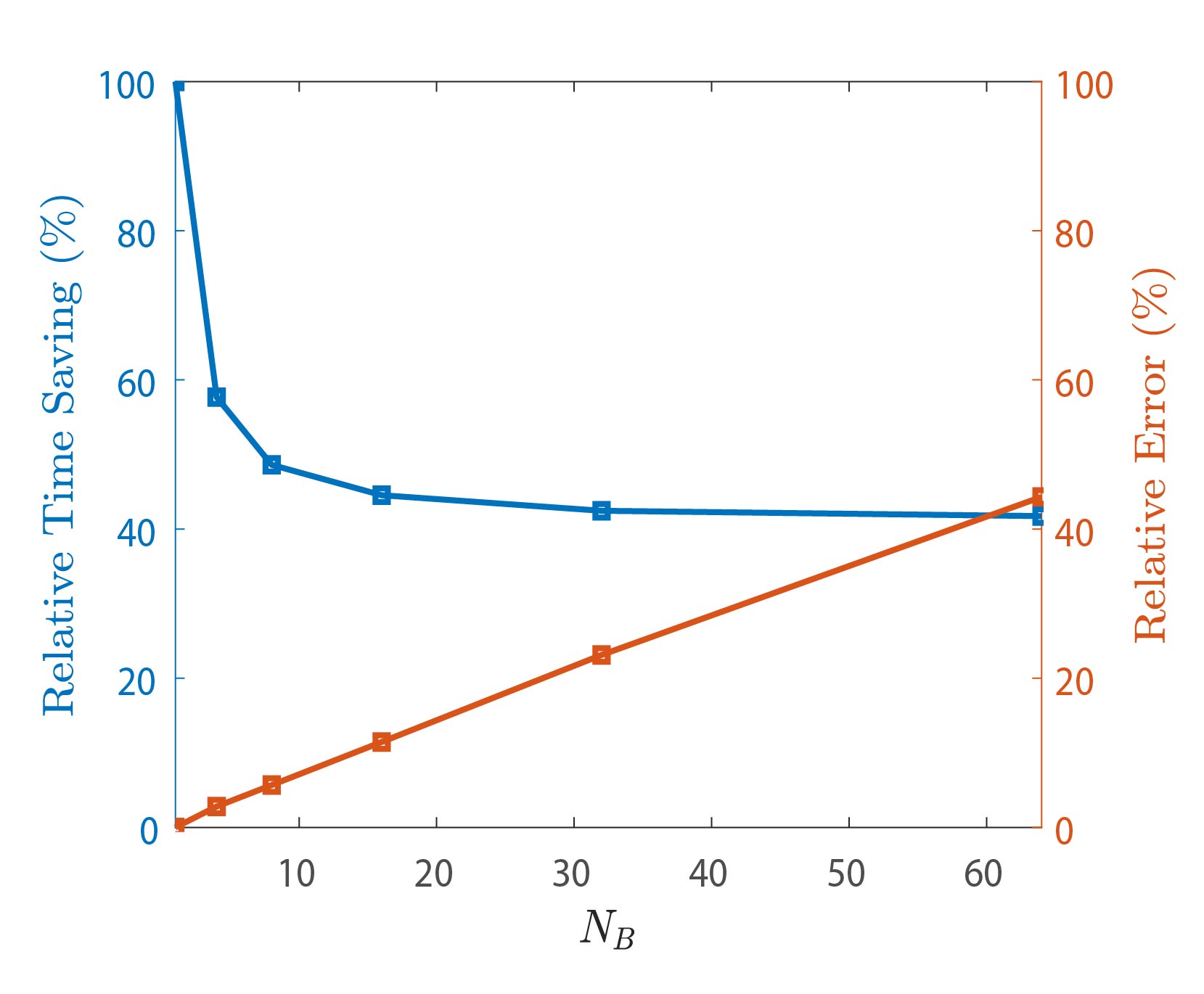}
	\caption{Plot of relative time saving (left $y$-axis) and relative error of reconstruction (right $y$-axis) vs slice binning factors ($N_B$).}
	\label{FigureResultsBinning_TimeSavingandError}
\end{figure}

To improve the reconstruction speed, recall that we combine $N_B$ consecutive 2D projected potentials to change the effective thickness of each layer. However, the reconstruction quality deteriorates as we gradually increase the number of slices being binned $N_B$. Therefore, the extent to which we can bin the slices is of special interest. The precise mathematical error analysis is not available due to the non-linearity of the multislice method, and so to estimate an upper bound for slice-burring we use the 3D  CTF of the imaging system by assuming single or weakly scattering~\cite{Tian:15:3D}. Then, we are able to linearize the problem to obtain an estimate of the error. In a traditional imaging system with numerical aperture $\mathrm{NA} = \lambda/\Delta x$, where $\Delta x$ is the pixel size, the axial resolution can be characterized as:
\begin{equation}
\Delta z = \lambda / (1 - \sqrt{1-\mathrm{NA}^2}). 
\end{equation}
Based on Nyquist sampling criterion, the maximum thickness for every slice should be less than $\Delta z$ to support the axial resolution at every angle. 

We test the effectiveness and fundamental limit of the proposed slice-binning method.  Here, we exponentially increase $N_B$ to examine the effect it has on reconstruction error, computation time, and convergence behavior of the algorithm. 

To simplify our discussion, all datasets in the validation process are generated from 60 uniformly separated tilt angles with 3 defocus planes, assuming infinite dose. We do not apply any regularization methods as they alter the convergence behavior depending on the choice of the regularization parameter.

\newpage
\bibliographystyle{elsarticle-num} 
\bibliography{ref}

\end{document}